\documentclass[journal=nalefd,manuscript=article]{achemso}
 \usepackage[utf8]{inputenc}
 \usepackage{multirow}
 \usepackage{caption}
 \usepackage{subcaption}
  \usepackage{comment}
  \usepackage{soul}
\usepackage[normalem]{ulem}
 \usepackage{amsmath,amssymb}
\usepackage{graphicx}
\usepackage{xcolor}
\author{Andrea Mucchietto}
\affiliation{\'Ecole Polytechnique F\'ed\'erale de Lausanne (EPFL), 1015 Lausanne, Switzerland, Laboratory of Nanoscale Magnetic Materials and Magnonics, Institute of Materials (IMX   )}
\author{Korbinian Baumgaertl}
\affiliation{\'Ecole Polytechnique F\'ed\'erale de Lausanne (EPFL), 1015 Lausanne, Switzerland, Laboratory of Nanoscale Magnetic Materials and Magnonics, Institute of Materials (IMX)}
\author{Dirk Grundler}
\email{dirk.grundler@epfl.ch}
\affiliation{\'Ecole Polytechnique F\'ed\'erale de Lausanne (EPFL), 1015 Lausanne, Switzerland, Laboratory of Nanoscale Magnetic Materials and Magnonics, Institute of Materials (IMX)}
\alsoaffiliation{'Ecole Polytechnique F\'ed\'erale de Lausanne (EPFL), 1015 Lausanne, Switzerland, Institute of Electrical and Micro Engineering (IEM)}
\title{{{Magnon-assisted magnetization reversal of Ni$_{\rm 81}$Fe$_{\rm 19}$ nanostripes on Y$_{\rm 3}$Fe$_{\rm 5}$O$_{\rm 12}$ with different interfaces}}}

\keywords{spin waves, magnetization reversal, YIG, broadband spectroscopy, magnetic interfaces}

\begin{document}
	
\begin{abstract}
	Magnetic bit writing by  {short-wave} magnons without conversion to the electrical domain is expected to be a game-changer for in-memory computing architectures. Recently, the reversal of nanomagnets by propagating magnons was demonstrated. However, experiments have not yet explored different wavelengths and \textcolor{blue}{the nonlinear excitation regime of magnons required for computational tasks}. We report on the magnetization reversal of individual 20-nm-thick Ni$_{81}$Fe$_{19}$ (Py) nanostripes integrated onto 113-nm-thick yttrium iron garnet (YIG). We suppress direct interlayer exchange coupling by an intermediate layer such as Cu and SiO$_2$. Exciting magnons in YIG with wavelengths $\lambda$ down to 148 nm we observe the reversal of the integrated ferromagnets in a small opposing field \textcolor{blue}{of 14 mT. Magnons with a small wavelength of $\lambda=195$~nm, i.e., twice the width of the Py nanostripes, induced the reversal at an unprecedentedly small spin-precessional power of about 1 nW after propagating over 15~$\mu$m in YIG. Considerations based on dynamic dipolar coupling explain the observed wavelength dependence of magnon-induced reversal efficiency. For an increased power the stripes reversed in an opposing field of only about 1 mT. Our findings are important for the practical implementation of nonvolatile storage of broadband magnon signals in YIG by means of bistable nanomagnets without the need of an appreciable global magnetic field.}
\end{abstract}

\maketitle

\newpage

\begin{figure}[htbp] 
	\centering
	\includegraphics[width=0.8\linewidth]{ 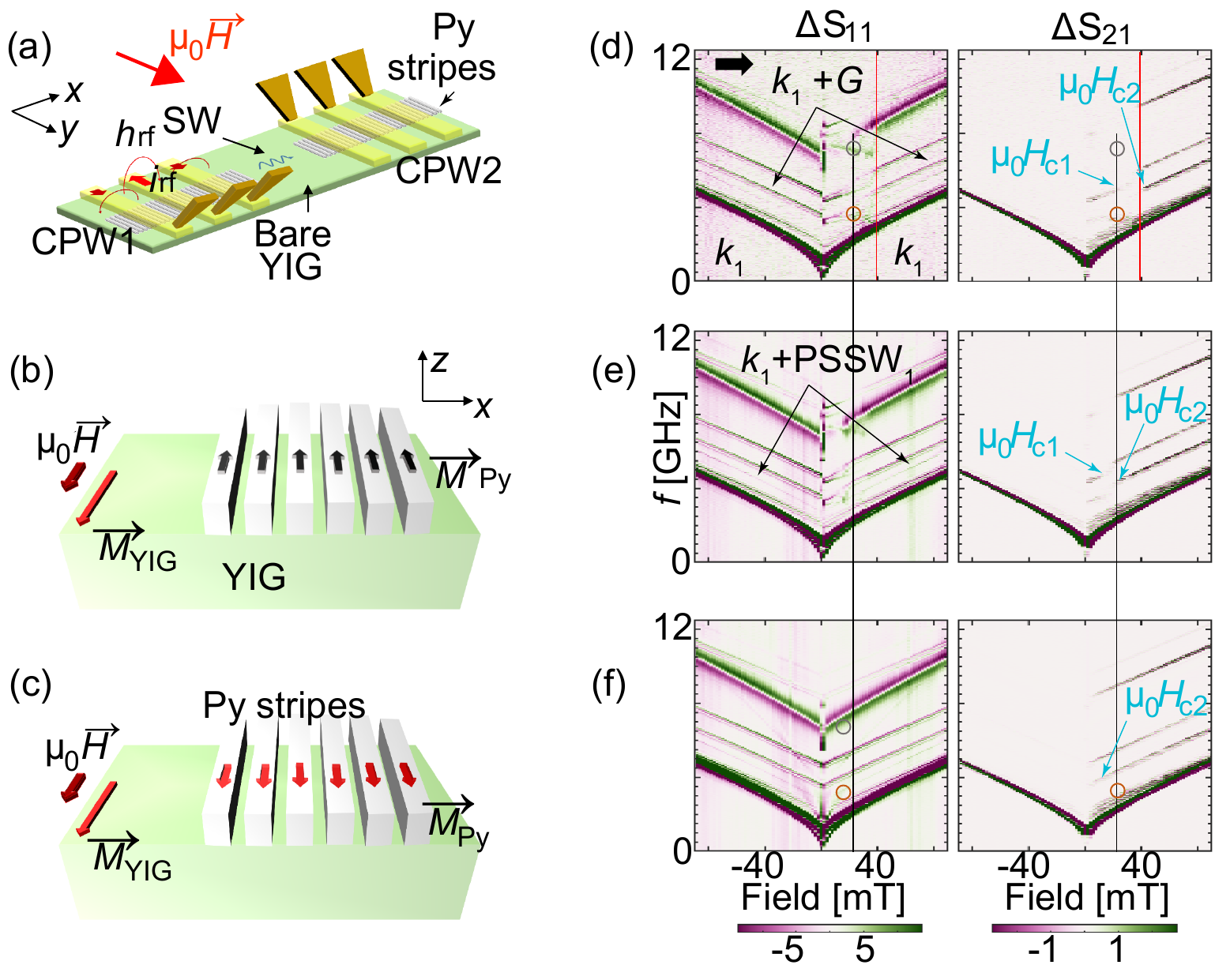}   
	\caption{(a) Schematic device with the two CPWs, Py stripes (gratings) and microwave tips connected to a VNA. A current $i_{\rm rf}$ at frequency $f_{\rm irr}$ is injected into CPW1. The generated field $h_{\rm rf}$ excites magnons. Sketches of the (b) anti-parallel (AP) and (c) parallel (P) magnetic configuration of Py nanostripes and YIG. Color-coded spectra $\Delta S_{11}$ (left) and $\Delta S_{21}$ (right) taken as a function of field from -90 mT to +90 mT {on sample \textcolor{blue}{A}} for powers $P_{\rm irr}$ of (d) $-30~$dBm, (e) $-15~$dBm, and (f) $0~$dBm. The horizontal arrow in (d) indicates the magnetic field sweep direction. We display $\Delta S$, i.e., the difference of scattering parameters $S$ that are taken at subsequent field values. Intense and dark colors indicate magnon resonances. Labels and symbols highlight specific resonances and critical fields. The black (red) vertical lines indicate 24 mT (40 mT). \textcolor{blue}{In (f, right) the AP branch is not resolved indicating $H_{C1}$ is (close to) zero.}  \label{cpw}}
\end{figure}

Collective spin excitations in a magnetically ordered material are called spin waves (SWs) or, in quantum-mechanical terms, magnons. By means of SWs, angular momentum is transferred without electrical charge motion, hence no Joule heating is generated. Therefore, SWs represent a new paradigm for signal processing at low power consumption and for a non-charge-based beyond-CMOS technology\cite{Khitun_2010,ChumakMagTrans,Mahmoud2020}. In magnonic applications, microwave signals are applied to integrated coplanar waveguides (CPWs) and excite coherent SWs in the  {adjacent} magnetic layer. Grating couplers consisting of ferromagnetic nanoelements [Fig. \ref{cpw}(a)] have proven to enhance the microwave-to-magnon coupling at GHz frequencies if integrated to CPWs \cite{Yu2013,yu2016approaching,PhysRevB.100.104427,KBaumgaertl2020}. They emit and detect magnons with wavelengths $\lambda$ down to below 50 nm in ferrimagnetic yttrium iron garnet (YIG) \cite{Liu2018,SWatanabe2023}. \textcolor{blue}{Wang et al. explored the spin wave emission from a ferromagnetic stripe into YIG \cite{ChiralPumpSW}. They explained the strong spin-wave signal in the underlying YIG by prominent dipole-dipole interaction without assuming spin currents.} Recently, it has been reported that dipolar SWs reversed \textcolor{blue}{100-nm-wide} ferromagnetic nanostripes deposited directly on YIG after propagating over 25 micrometers. \cite{baumgaertl2023reversal} The magnon-induced switching of Ni$_{81}$Fe$_{19}$ (Py) nanostripes on YIG occurred in the linear excitation regime at low microwave power. However, the wavelength $\lambda$ used for switching remote nanostripes was a few micrometers long. Such value of $\lambda$ is not adequate for nanomagnonic in-memory computing in either the linear or nonlinear excitation regime \cite{Islam_2019,Sebastian2020,Papp2021}.
\\ \indent
In this work we report \textcolor{blue}{remote switching of 100-nm-wide Py nanostripes by magnons with $\lambda$ down to 148 nm in YIG. We explore different interfaces and, both, the linear and nonlinear excitation regime. The Py nanostripes were integrated on an intermediate layer of either Cu or SiO$_2$ on YIG. Thereby we suppressed the direct exchange coupling \cite{Klingler2018} between Py and YIG. Using the identical nanostripe design, we compare our results to Ref. \cite{baumgaertl2023reversal} in which an intermediate layer between Py and YIG was avoided.} Using broadband spectroscopy [Fig. \ref{cpw}(d) to (f)] and spatially resolved Brillouin light scattering (BLS) we acquire magnon spectra before and after exciting propagating magnons of different $\lambda$ in YIG. \textcolor{blue}{We observe irreversible changes in BLS spectra which indicate reversed states of Py magnetization vectors $\mathbf{M}_{\rm Py}$ which we attribute to magnon-induced switching in a small opposing field. We analyze the power absorbed by the precessing spins in YIG and find that propagating magnons whose wavelength is twice the nanostripe width show the minimum power level of \textcolor{blue}{about 1} nW representing the highest reversal efficiency. Our findings go beyond earlier reports in that we (i) demonstrate experimentally that dynamic dipolar coupling between Py and YIG is sufficient for magnon-induced reversal, (ii) explain the wavelength dependent reversal efficiency, and (iii) report switching by propagating magnons with a wavelength of only 148 nm. (iv) Our BLS data reveal the magnon-induced reversal in the non-linear excitation regime which we attribute to parametrically pumped magnons. Our findings pave the way for future in-memory computation in linear and non-linear nanomagnonics with materials combinations which do not require direct exchange coupling between the magnetic elements.}\\

\section*{Results}
\begin{figure}[ht]
	\centering
	\includegraphics[width=0.85\linewidth]{ 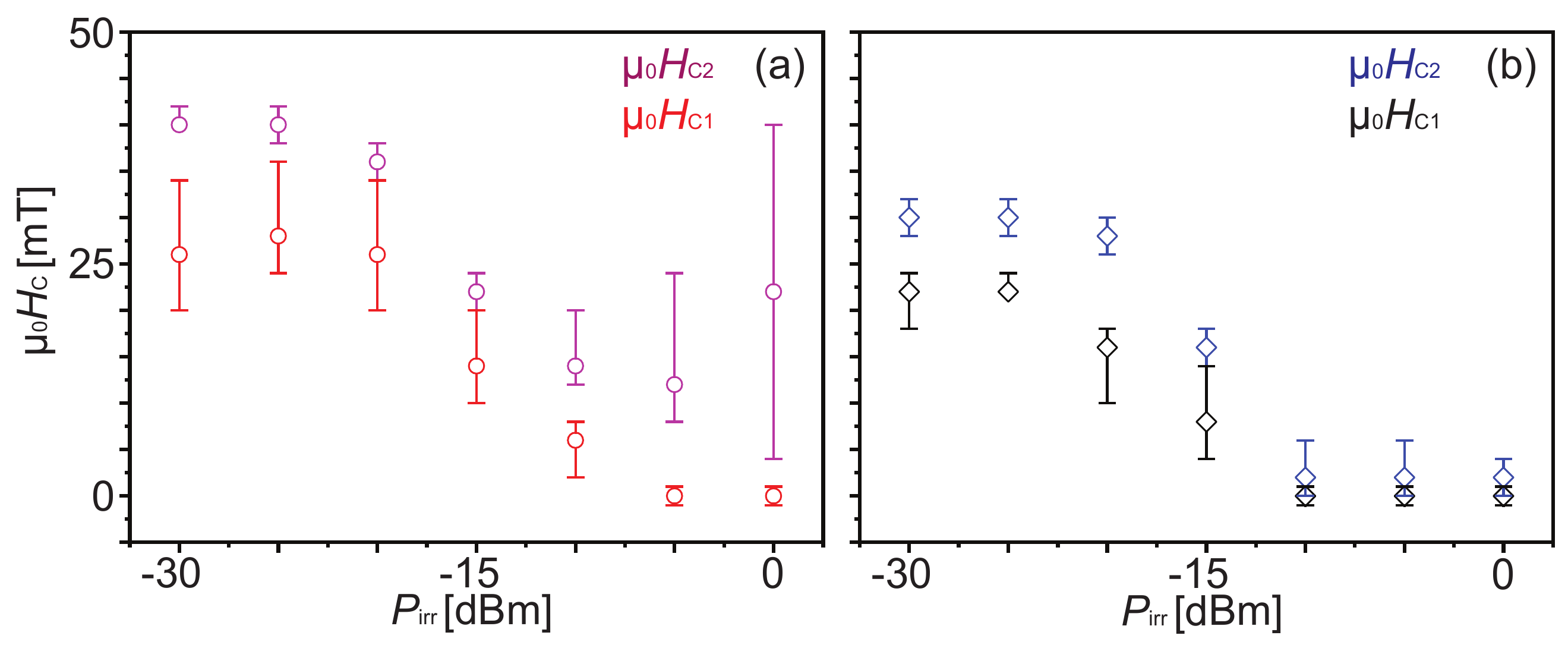}
\caption{ {{For samples (a) \textcolor{blue}{A} and  (b) B the critical fields $\mu_0H_{\rm C1}$ and $\mu_0H_{\rm C2}$ realizing 50\% of the maximum signal strengths of the two relevant magnon branches are shown as a function of $P_{\rm irr}$. The error bar refers to fields needed to achieve 30\% and 70\% of the maximum signal strengths of branches AP and P.}}\label{switchingfield}}
\end{figure}

We fabricated one-dimensional (1D) periodic arrays of Py nanostripes (gratings) on 113-nm-thick YIG which was commercially available from the same supplier as in Refs. \cite{KBaumgaertl2020,Maendl2018,Watanabe2021}. 
\textcolor{blue}{The stripes consisted of 20-nm-thick Py and were 100 nm wide. They are arranged with a period of $p=200~$ nm. The stripe lengths were consistent with {Ref. \cite{baumgaertl2023reversal}} and alternated between 25 and 27 $\mu$m. The total width of a grating amounted to $w_{\rm GC} = 10~\mu$m. They were fabricated on YIG with a 5-nm-thick intermediate layer of either Cu (sample A) or SiO$_2$ (sample B).} We introduced the intermediate layers to intentionally modify the coupling between Py and YIG compared to 
\textcolor{blue}{Ref. \cite{baumgaertl2023reversal} where Py had been deposited directly on YIG.} \textcolor{blue}{The intermediate layers suppress the exchange coupling. Moreover, in sample B the SiO$_2$ spacer avoids the spin pumping mechanism thus allowing for only dipolar coupling between Py and YIG. In sample A, the Cu spacer thickness is smaller than the spin diffusion length \cite{SpinDiffLengthRev} and a spin pumping related torque could occur in addition to dipolar coupling \cite{Suresh2021}. In the following we denote the sample without an intermediate layer used in Ref. \cite{baumgaertl2023reversal} as sample C.} In the coplanar waveguides (CPWs)  [Fig. \ref{cpw}(a)], the Au lines (gaps) were 2.1 $\mu$m (1.4 $\mu$m) wide. The distance between signal lines of two parallel CPWs was 15 $\mu$m.  A finite element analysis using COMSOL Multiphysics provided the inhomogeneous radiofrequency (rf) field $h_{\rm rf}$ of the CPWs (Fig. S1). Without the lattice of nanostripes, they excited and detected most efficiently spin waves with a wave vector $k$ of $k_1= 0.87~{\rm rad}/\mu$m. An in-plane magnetic field $\mathbf{H}$ was applied to realize specific magnetic histories and controlled different relative orientations of magnetization vectors in Py ($\mathbf{M}_{\rm Py}$) and YIG ($\mathbf{M}_{\rm YIG}$) [Fig. \ref{cpw}(b) and (c)]. We performed broadband measurements (Methods) of scattering parameters $\Delta S_{11}$ (reflection) and $\Delta S_{21}$ (transmission) with port 1 and port 2 of a vector network analyzer (VNA) connected to CPW1 and CPW2, respectively. We observed several resonant branches above the $k_1$ excitation in Fig. \ref{cpw}(d) to (f). Considering Refs. \cite{KBaumgaertl2020,yu2016approaching,Liu2018}, the additional high-frequency branches reflected grating coupler modes such as $k_1+G$, with $G=2\pi/p$, different orders of perpendicular standing spin waves (PSSWs), and the magnetic resonance in the Py nanostripes. The latter one was the prominent high-frequency branch in $\rm \Delta $$S_{\rm 11}$ which started at about 10.5 GHz at -90 mT in Fig. \ref{cpw}(d). \textcolor{blue}{In Figs. S2 and S3 we report further spectra from which we extracted the quasi-static characteristics of samples A and B}. We applied BLS microscopy ($\mu$BLS) in that we focussed laser light for inelastic light scattering on Py nanostripes in the different gaps of a CPW. The laser spot diameter was about 400 nm. We note that the resonance frequency of Py nanostripes  was high (low) if $\mathbf{M}_{\rm Py}$ was parallel (anti-parallel) to the applied field $\mathbf{H}$ (see below). \cite{Gurevich96} \textcolor{blue}{The BLS microscopy was used to gain spatially resolved information about Py nanostripe reversal and explore the non-linear regime which was not achieved in Ref. \cite{baumgaertl2023reversal}.}\\ \indent
For obtaining the spectra $\Delta S_{11}$ and $\Delta S_{21}$ of sample \textcolor{blue}{A} in Fig. \ref{cpw}(d) to (f) we applied  $\mathbf{H}$ along the $y$-direction of {sample C}. We measured the scattering parameters in the following order: $S_{11}$, $S_{21}$, $S_{22}$ and $S_{12}$. \textcolor{blue}{In the following we focus on spin waves which were excited at CPW1 and propagated to CPW2, i.e., we report spectra $S_{11}$ and $S_{21}$, respectively. The spectra $S_{22}$ and $S_{12}$ showed consistent features when considering nonreciprocity and applying an inverted magnetic history.} We varied $\mu_0H$ from -90 mT to +90 mT [indicated by the black horizontal arrow in Fig. \ref{cpw}(d)] in steps of 2 mT. The nonreciprocal spin wave characteristics led to the large signal-to-noise ratios at positive $H$. The same measurement protocol was repeated for different VNA powers $P_{\rm irr}= -30, -15$ and $0$~dBm in Fig. \ref{cpw} (from top to bottom). At small power, we interpreted the branches of Fig. \ref{cpw}(d) such that at small positive $\mu_0H$ below \textcolor{blue}{$\mu_0 H_{\rm C1}=26$ mT [Fig. \ref{switchingfield}(a)]} the magnetization vectors of YIG and Py nanostripes were anti-parallel (AP) [Fig. \ref{cpw}(b)], in agreement with Co nanostripes on YIG reported in Ref. \cite{Liu2018}. In this field regime, the branch with negative slope $df/dH$ in $\Delta S_{11}$ [marked by a grey circle in Fig. \ref{cpw}(d)] was attributed to the ferromagnetic resonance inside the Py nanostripes. Their magnetization vectors $\mathbf{M}_{\rm Py}$ pointed still in $-y$-direction and against the applied positive field $\mathbf{H}$. They were anti-parallel also with $\mathbf{M}_{\rm YIG}$ as YIG had a coercive field $\leq 2~$mT. At small applied power $P_{\rm irr}=-30~$dBm, several of the grating coupler modes gained abruptly a pronounced signal strength at 40 mT (indicated by the red dashed line). We attributed this observation to the critical field $\mu_0H_{\rm C2}$ \textcolor{blue}{[Fig. \ref{switchingfield}(a)]} at which the reversal of the Py nanostripes underneath CPW2 (i.e., the detector CPW) occurred. For $\mu_0H>40~$mT, all the detected branches in Fig. \ref{cpw}(d) were similar to the ones at the correspondingly large negative fields. These branches indicated that the magnetization vectors of Py nanostripe lattices underneath both CPWs were now parallel (P) with $\mathbf{H}$ and $\mathbf{M}_{\rm YIG}$. Correspondingly, the transmission data showed the richest spectra of grating coupler modes [Fig. \ref{cpw}(d) on the right]. \\ \indent
For the spectra $\Delta S_{11}$ shown in Fig. \ref{cpw}(e), we used a larger power $P_{\rm irr}$ of -15 dBm. In the transmission spectra $\Delta S_{21}$ [right panel in Fig. \ref{cpw}(e)], \textcolor{blue}{the AP branch ended at a smaller field value} \textcolor{blue}{$\mu_0 H_{\rm C1}=14$~mT} and the region P started near \textcolor{blue}{$\mu_0H_{\rm C2}=22~$mT} instead of 40 mT. The Py nanostripes underneath CPW1 and CPW2, respectively, experienced a smaller field region of anti-parallel alignment with $\mathbf{M}_{\rm YIG}$ compared to Fig. \ref{cpw}(d). This observation indicated that the larger VNA power $P_{\rm irr}$ used for broadband spectroscopy led to the reversal of Py nanostripes underneath both CPW1 and CPW2.\\ \indent The onset of the P region occurred at an even smaller $H$ in Fig. \ref{cpw}(f) when $P_{\rm irr}$ of 0 dBm was used. The branches attributed to grating coupler modes in the P region {showed a weak signal strength \textcolor{blue}{already} at $\mu_0H=2~$mT which increased with increasing $H$. This means that close to zero field the magnetization vector $\mathbf{M}_{\rm Py}$ underneath CPW1 pointed into the $+y$-direction \textcolor{blue}{(P configuration)}, i.e., $\mu_0H_{\rm C1}\leq 2~$mT.} The reversal field of the Py nanostripes under CPW1  \textcolor{blue}{of sample A} was hence reduced by about  {26 mT} when applying $P_{\rm irr}=0$~dBm (1 mW) compared to $P_{\rm irr}=-30~$dBm (1 $\mu$W). Such a large reduction of {$\mu_0H_{\rm C1}$ was not reported in Ref. \cite{baumgaertl2023reversal} (sample \textcolor{blue}{C}) which had Py directly deposited on YIG.}\\
\indent {To characterize the power-dependent switching field distribution for samples A and B (Fig. \ref{switchingfield}) we adopted the methodology developed in Ref. \cite{baumgaertl2023reversal}. We evaluated VNA spectra taken at many different powers $P_{\rm irr}$ and analyzed the field-dependent signal strengths of the first GC branch in the AP and P state. {In such experiments, we applied $i_{\rm rf}$ covering a broad frequency regime from about 10 MHz to 20 GHz. Assuming that the magnon mode for most efficient switching resided in this frequency regime, we obtained the minimum critical field values for reversal at a given $P_{\rm irr}$.}  \textcolor{blue}{At each value of $P_{\rm irr}$, we extracted the critical field values $\mu_{\rm 0}H_{\rm C1}$ and $\mu_{\rm 0}H_{\rm C2}$ that corresponded to 50\% of the maximum signal strengths of the AP and P branches, respectively (symbols in Fig. \ref{switchingfield}).} The difference ($H_{\rm C2}-H_{\rm C1}$) reflected the distribution of switching fields of nominally identical Py nanostripes underneath the two CPWs. \textcolor{blue}{In sample A (Fig. \ref{switchingfield}a), the switching fields were distributed over a larger field range} than in sample B (Fig. \ref{switchingfield}b) for $P_{\rm irr}<-20~$dBm.}\\ \indent
{We first consider the critical fields {$\mu_{\rm 0}H_{\rm C1}$ extracted from the AP branches of both samples \textcolor{blue}{A} and B. At low power, $P_{\rm irr}\, \leq$ -25 dBm, {$\mu_{\rm 0}H_{\rm C1}$ is comparable within error bars in both samples. $\mu_{\rm 0}H_{\rm C1}$ decreases to $0 \, (\pm 1)~$mT in sample \textcolor{blue}{A} (B) when $P_{\rm irr} \, \geq$ -5 dBm (-10 dBm). We now focus on the critical fields of the P branch, i.e. $\mu_{\rm 0}H_{\rm C2}$. For $P_{\rm irr}<-20~$dBm the critical field $\mu_{\rm 0}H_{\rm C2}$ is larger for sample \textcolor{blue}{A} than for sample B (cf. Fig. \ref{switchingfield}a and \ref{switchingfield}b). $\mu_{\rm 0}H_{\rm C2}$ decreases as $P_{\rm irr}$ increases up to -5 dBm. $\mu_{\rm 0}H_{\rm C2}$ is reduced to 2 mT in sample B when $P_{\rm irr} \, \geq$ -10 dBm and it maintains this small value at larger $P_{\rm irr}$. This is the smallest value so far detected for reversal of Py nanostripes on YIG induced by propagating magnons. \textcolor{blue}{This finding is one of the key achievements of this work. Near-zero critical fields were not be observed in Ref. }\cite{baumgaertl2023reversal}\\ \indent For comparison, in sample \textcolor{blue}{A}, the critical field $\mu_{\rm 0}H_{\rm C2}$ is about 15 mT at $-5~$dBm. At the largest $P_{\rm irr}$, it has increased again (Fig. \ref{switchingfield}a) and reaches $\approx$ 22 mT. Considering Ref. \cite{baumgaertl2023reversal} \textcolor{blue}{we attribute this increase in critical fields of Py nanostripes underneath CPW2 at high $P_{\rm irr}$ to the nonlinear regime of magnon excitation underneath CPW1 with enhanced magnon scattering. Because of the scattering processes the magnon amplitudes after 15~$\mu$m are below the threshold for complete reversal of the nanostripe array under CPW2.}
The incomplete reversal at high power is observed for both samples A and C where spin pumping is allowed. In sample B (Fig. \ref{switchingfield}b) we do not observe an increase in critical fields at large powers. Instead, we find the largest reduction in critical fields $H_{\rm C1}$ and $H_{\rm C2}$ in sample B. Here, a 5-nm-thick SiO$_{\rm 2}$ spacer rules out that spin pumping is relevant for the efficient magnon-induced reversal. \textcolor{blue}{We note that the insertion of both the SiO$_{\rm 2}$ and Cu spacer excludes the transfer of exchange magnons which was assumed in Refs. \cite{Han1121,Wang2019b,Guo2021,Zheng2022}. Our experiments highlight the importance of dynamic dipolar coupling between Py and YIG when developing a microscopic understanding of the magnon-induced reversal mechanism.}
\\
\indent\begin{figure}[ht]
	\centering
	\includegraphics[width=0.9\linewidth]{ 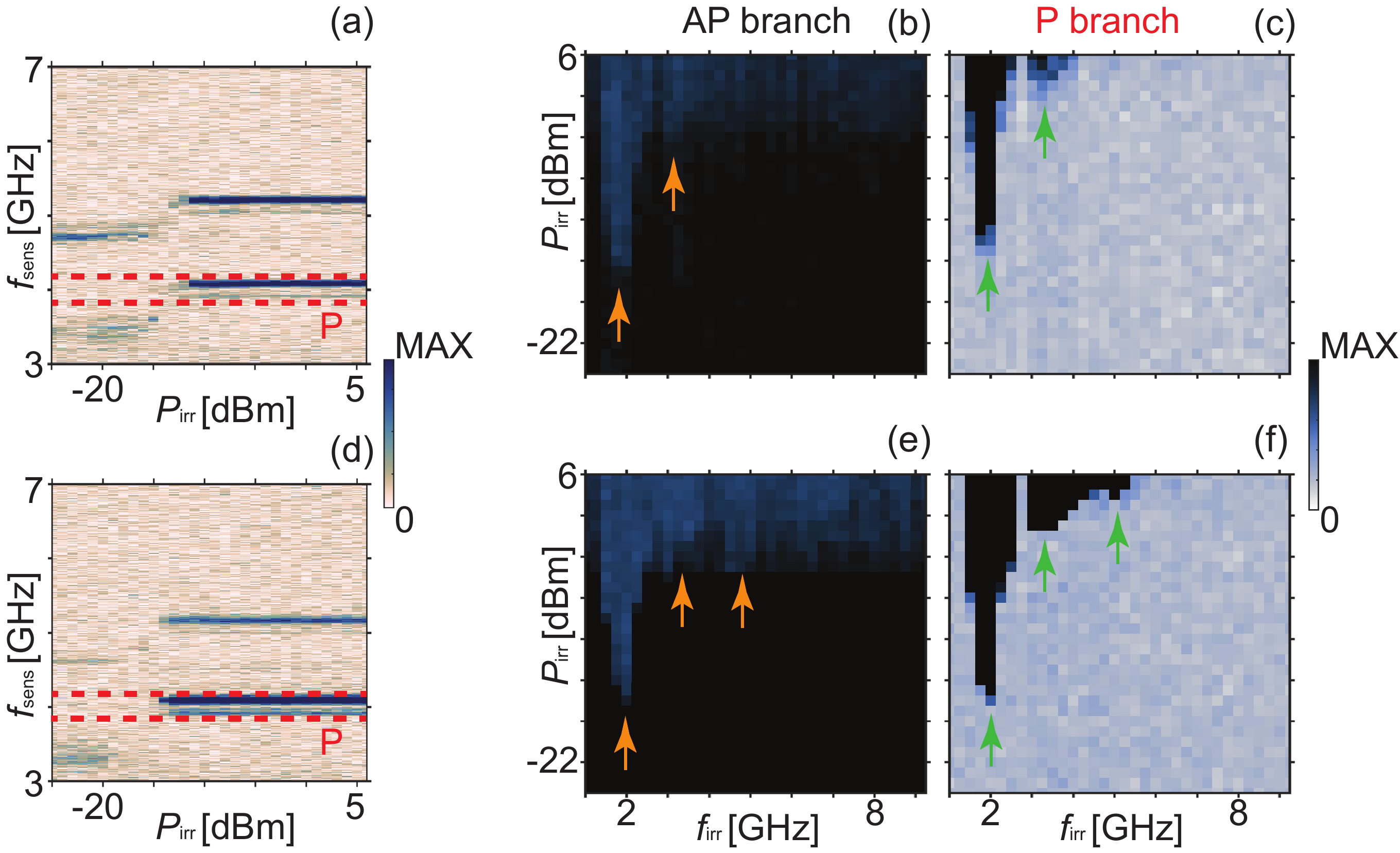}
	\caption{(a) Mag($S_{\rm 21}$) recorded on  {sample C (Cu spacer) between $f_{\rm sens}=3$ and $7$~GHz at +14~mT with a power of -25~dBm after applying a microwave signal with $f_{\rm irr}=1.75~$GHz to CPW1 for increasing power $P_{\rm irr}$. Switching yield maps at +14~mT for sample C displaying color-coded (b) Mag($S_{\rm 11}$) and (c) Mag($S_{\rm 21}$) integrated as a function of $f_{\rm sens}$ for the AP and P branch respectively. The frequency integration range for the P branch is highlighted by the red dashed lines in (a). To extract the switching yield map for the AP branch, the first GC mode branch in the Mag($S_{\rm 11}$) spectrum is used. Panels (d) to (f) show the corresponding dataset for sample B (SiO$_{\rm 2}$ spacer). Arrows indicate local minima in the power threshold inducing stripe reversal by specific magnons discussed in the text.} \label{switchyield}}
\end{figure}
To quantify the power level at which a specific spin wave mode in YIG reversed nanostripes we followed the concept of switching yield maps (Methods) introduced in Ref. \cite{baumgaertl2023reversal} (Fig.~\ref{switchyield}). The samples were first saturated at -90 mT applied along the $y$-axis. Then, the field was gradually increased to +14 mT and kept constant. We provided powers ${P}_{\rm irr}$ ranging from -25 to +6 dBm with +1 dBm steps within a 0.25-GHz-wide frequency window starting at a specific frequency \textit{f}$_{\rm irr}$. After each power step and corresponding irradiation for 1 msec, the VNA power level was reduced to $-25~$dBm and the transmitted signal ($S_{21}$) was recorded as a function of frequency $f_{\rm sens}$ ranging from 3 to 7 GHz. Figure ~\ref{switchyield} displays such datasets in panels (a) and (d) as well as gray-scaled switching yield maps performed at +14 mT for sample \textcolor{blue}{A} (top row) and sample B (bottom row). The maps labelled by AP (P) branch \textcolor{blue}{in Fig. \ref{switchyield}b and e (Fig. \ref{switchyield}c and f) reflect the reversal of Py nanostripes underneath CPW1 (CPW2) of sample A and B, respectively}.
From these maps, we extracted the critical power levels for magnon-assisted switching at CPW1 and CPW2 which we denote by $P_{\rm C1}$ and $P_{\rm C2}$, respectively.  \textcolor{blue}{In Fig. \ref{criticalpowerstabrecap}, we particularly display the critical power values extracted near the local minima indicated by arrows in Fig. \ref{switchyield} reflecting modes $k_1$, $k_1+G$ and ${k}_{\rm 1} + {G} + {k}_{\rm PSSW1}$ (from left to right)}.
\begin{figure}[h!]
	\centering
	\includegraphics[width=0.5\linewidth]{ 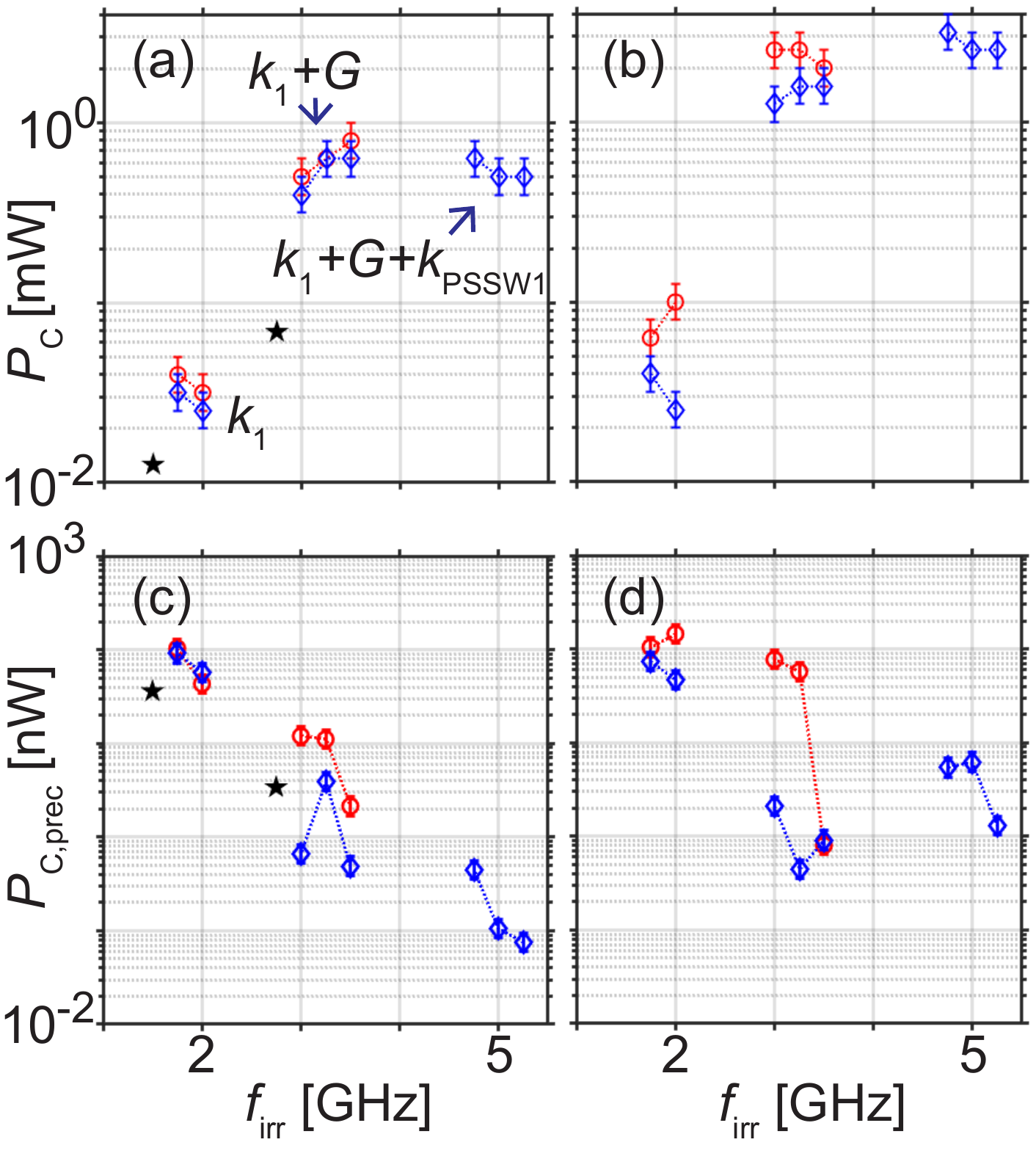}
	\caption{For samples \textcolor{blue}{A} (red), B (blue), and C (black, magenta) the critical powers (a) $P_{\rm C1}$ and (b) $P_{\rm C2}$, (c) $P_{\rm C1,prec}$ and (d) $P_{\rm C2,prec}$ are depicted for irradiation frequencies $f_{\rm irr}$ in half-logarithmic graphs. In (a) we label magnon modes by $k_{\rm 1}$, $k_{\rm 1} + G$ and $k_{\rm 1} + G + k_{\rm PSSW1}$. In (d) the values of sample C (magenta symbols) are scaled by a factor $\rho$ to correct for the different propagation path length and decay of magnon amplitudes between CPWs (Methods, paragraph C). Connecting lines are guide to the eyes.\label{criticalpowerstabrecap}}
\end{figure}
When exciting the $k_1$ mode near 2~GHz in sample \textcolor{blue}{A} and B at $+14~$mT, we require $P_{\rm C1}$ between \textcolor{blue}{30 and 40} $\mu$W for the reversal of 50\% of the nanostripes below CPW1 [red symbols in Fig. \ref{criticalpowerstabrecap}(a)]. This power value is only about a factor of three larger than the one of sample C published in Ref. \cite{baumgaertl2023reversal} [black symbol near 1.5 GHz in Fig. \ref{criticalpowerstabrecap}(a)].
For all samples, $P_{\rm C1}$ and $P_{\rm C2}$ increase with increasing mode frequency. For the reversal underneath CPW1 by means of the GC mode $k_1+G$ (excited between 2.75 and 3.5 GHz) VNA powers $P_{\rm C1}$ of 400 to 800 $\mu$W are required. The reversal of Py nanostripes underneath CPW2 is achieved at a further increased power level $P_{\rm C2}$ of up to 2.5 mW. For sample C, $P_{\rm C2}$ was larger than 3 mW and not determined. \\ \indent \textcolor{blue}{To compare different samples, it is instructive to consider the power values $P_{\rm C1, prec}$ [Fig. \ref{criticalpowerstabrecap}(c)] and $P_{\rm C2, prec}$ [Fig. \ref{criticalpowerstabrecap}(d)] which quantify the power absorbed by the spin-precessional (prec) motion in YIG at the emitter CPW (Methods). These values consider that only part of the rf power applied by the VNA is absorbed by the spin system and converted into magnons. These values, taken at the same field, allow us to compare the different samples independent of the individual efficiency of microwave-to-magnon transduction.} In case of the \textcolor{blue}{long-wavelength} modes $k_1$ existing near 2 GHz in samples A and B, we observe reversal at power levels $P_{\rm C1, prec}$ between 4 to 10 nW [Fig. \ref{criticalpowerstabrecap}(c)]. $P_{\rm C2, prec}$ for modes $k_1$ is only slightly larger attributed to a weak decay of the magnon mode between emitter and detector CPW. 
At larger excitation frequencies $f_{\rm irr}$ between 3 and 3.5 GHz corresponding to the first GC mode resonance $k_1+G$, power values $P_{\rm C1,prec}$ are smaller by up to two orders of magnitude compared to modes $k_1$ in Fig. \ref{criticalpowerstabrecap}(c). Here, sample \textcolor{blue}{B} realizes the smallest values $P_{\rm C1,prec}$ down to about \textcolor{blue}{ 0.5} nW. 
Note that despite larger coercivities in sample \textcolor{blue}{A} the magnon-induced reversal underneath CPW1 via mode $k_1+G$ is realized at a smaller power than in sample C with the direct interface between Py and YIG. \textcolor{blue}{A similar small value of about 1 nW is found in Fig. \ref{criticalpowerstabrecap}(d) for reversal underneath CPW2, suggesting a weak decay of magnon amplitudes after a path of 15~$\mu$m. The key finding of Fig. \ref{criticalpowerstabrecap}(d) is} that the mode $k_1+1G$ in sample B is most efficient in terms of $P_{\rm C2,prec}$ and nanostripe reversal underneath CPW2. Considering its intermediate layer to be an insulator (SiO$_2$) the dipolar coupling between magnons in YIG and Py provides the torque for the reversal. 

We note that in sample B we observe nanostripe reversal by a further mode with wavelength {$\lambda = 148~$nm} corresponding to a wavevector $k=\sqrt{(\mathbf{k}_{\rm 1} + \mathbf{G})^2 + k^2_{\rm PSSW1}}$ ($k_{\rm PSSW1}$ is the first quantized magnon mode across the YIG thickness). \textcolor{blue}{When exciting this mode in the frequency range 4.75 to 5.25 GHz, we extract $P_{\rm C2,prec} = 1.3$ to 5.4 nW. These power values are increased compared to $P_{\rm C2,prec}$ found near 3 GHz. For sample B, we observe the smallest spin-precessional power $P_{\rm C1,prec}$ for reversal in Fig. \ref{criticalpowerstabrecap}(c) at 5.25 GHz. We explain the small value by the combined effect of magnon-induced reversal and microwave-assisted switching near the eigenresonance of the Py nanostripes underneath CPW1 before their reversal at +14 mT. For the nanostripes underneath CPW2 ($P_{\rm C2,prec}$) the microwave-assisted switching does not play a role due their large separation from the CPW1 attached to the rf source.}\\ \indent
\begin{figure}[ht] 
	\centering
	\includegraphics[width=0.75\columnwidth]{ 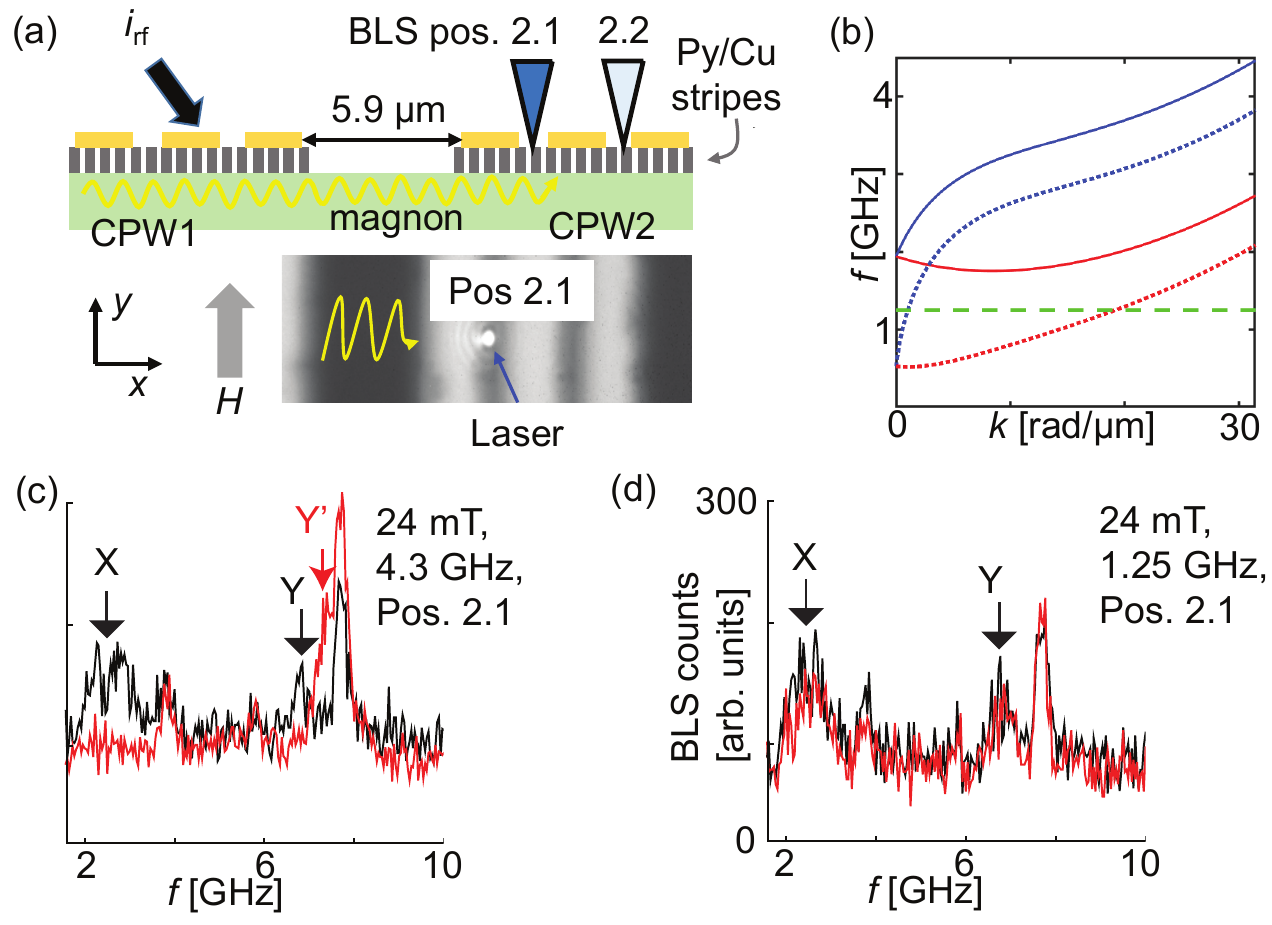}
	\caption{(a) Sketched cross-section of the device (top). The CPW lines (yellow) are on top of the stripes (dark grey) which have been fabricated on YIG (green). In BLS we detected thermally excited magnons in Pos. 2.1 (microscopy image) and 2.2. (b) Magnon dispersions in the thin YIG at 24 (solid lines) and 2 (dotted lines) mT calculated via the Kalinikos-Slavin formalism for two limiting configurations, i.e. Damon-Eshbach (blue lines) and backward volume (red lines) configuration. The horizontal dashed green line indicates 1.25 GHz which is below (inside) the magnon band at 24 mT (2 mT). Magnon spectra in Pos 2.1 at 24 mT before (black) and after (red) applying $i_{\rm rf}$ to CPW1 {for $P_{\rm irr}$ = 16 dBm} with (c) 4.3 GHz and (d)  1.25 GHz. Labels X, Y and Y' indicate characteristic resonant modes.\label{BLSoverview}}
\end{figure}
In the following we apply micro-focus BLS to sample \textcolor{blue}{A} [Fig.~\ref{BLSoverview}(a)] and gain spatially resolved information about the magnon modes that modify the magnetization vectors $\mathbf{M}_{\rm Py}$ of Py stripes. We do not evaluate absolute power values here as the BLS setup has not allowed for calibration, and CPWs were wire-bonded.
\begin{figure}[h!]
	\centering
	\includegraphics[width=0.75\linewidth]{ 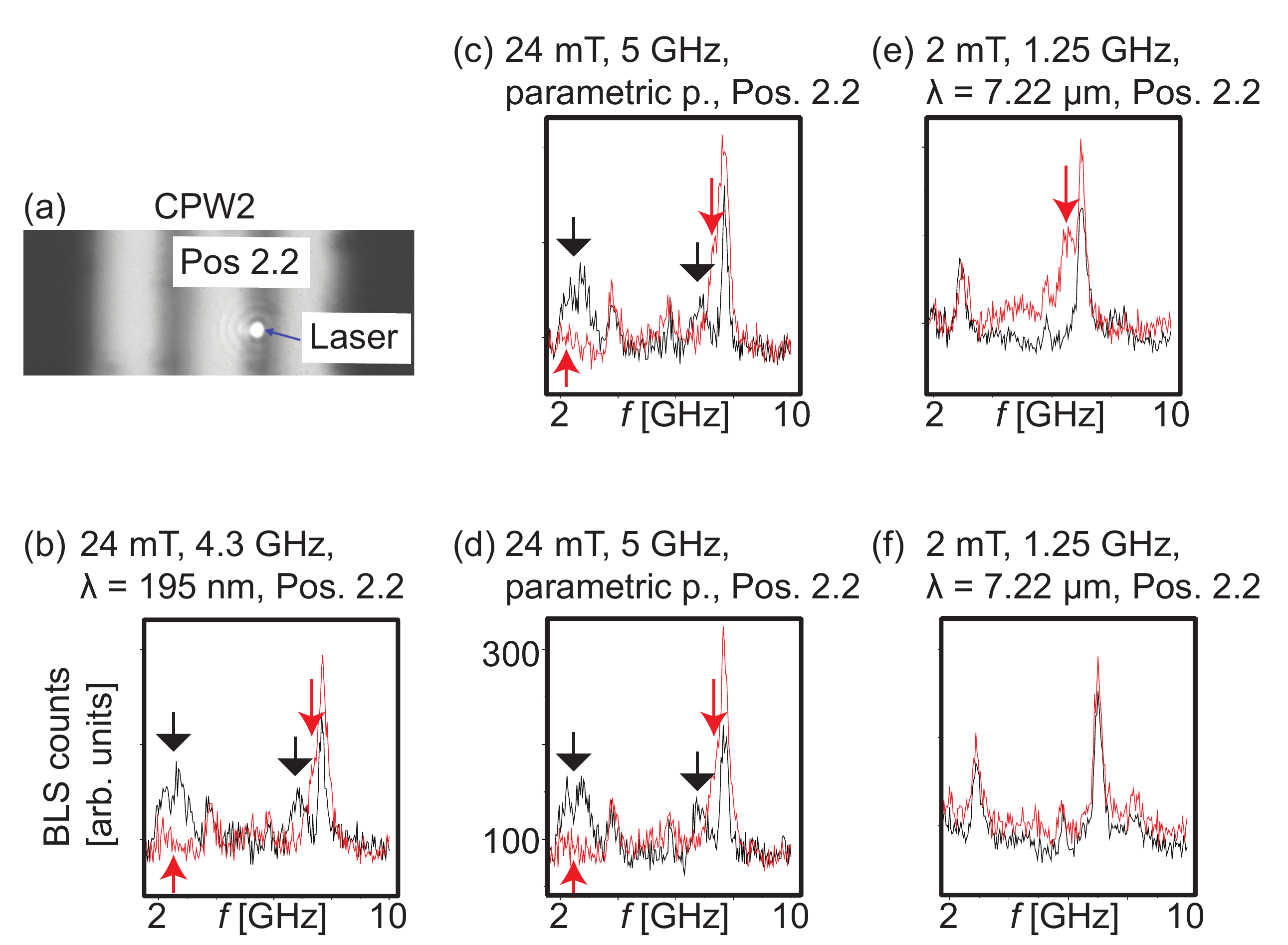}
	\caption{(a) Optical image when positioning the laser at Pos. 2.2. (b) Magnon spectra taken in Pos. 2.2 at 24 mT before (black) and after (red) applying $i_{\rm rf}$ to CPW1 with $f_{\rm irr}=4.3~$GHz. The reversal of Py nanostripes by magnons $k_1+G$ is evidenced. Magnon spectra in (c) Pos. 2.1 and (d) Pos. 2.2 before (black) and after (red) applying $i_{\rm rf}$ with $f_{\rm irr}=5~$GHz at CPW1. The black (red) arrows highlight characteristic modes (changes). Thermal magnon spectra in (e) Pos. 2.1 and (f) Pos. 2.2 before (black) and after (red) emitting magnons with $k_1$ by applying $i_{\rm rf}$ with $f_{\rm irr}=1.25~$GHz at CPW1. {In all these experiments the irradiation power was $P_{\rm irr}$ = 16 dBm}.  The legends list relevant parameters and highlight the parametric pumping (p.) experiments.\label{1g25}}
\end{figure}
We discuss BLS spectra reflecting the incoherent magnons excited thermally at room temperature. \textcolor{blue}{We compare spectra taken} before (black curves) and after (red curves) applying microwaves to CPW1. We explore different fields $H$ modifying the spin-wave dispersion relation in YIG [Fig.~\ref{BLSoverview}(b)] and different $f_{\rm irr}$. The laser wavelength (power) was 473 nm ({0.8} mW). Given the laser spot diameter of about 400~nm, we collected the Stokes's signal of magnons from up to two Py nanostripes and the underlying YIG. Each spectrum in Fig. ~\ref{BLSoverview}(c) and (d) had an acquisition time of approximately 2 hours.\\ \indent The spectrum shown as the black curve in Fig. \ref{BLSoverview}(c) displays magnon resonances existing in the gap of CPW2 in Pos. 2.1 for $\mu_0H=24$~mT after saturation along $-y$-direction using $\mu_0H=-84$~mT and before applying $i_{\rm rf}$ to CPW1. The frequencies of resonances marked X and Y indicate that the Py nanostripes are anti-parallel to $\mathbf{H}$ [Fig. \ref{cpw}(b)]. They are consistent with the frequencies marked by brown and grey circles, respectively, in Fig. \ref{cpw}(d). After applying $i_{\rm rf}$ at $f_{\rm irr}=4.3~$GHz to CPW1 with a nominal irradiation power $P_{\rm irr}$ of up to {39.8} mW (16 dBm) the red spectrum was obtained at the same position. Due to wire-bonded connections we expected the power in CPW1 to be a few dB lower than the nominal value. The red spectrum is markedly modified compared to the black curve: the resonance peaks X and Y reduced to the noise level, and a higher frequency resonance Y' was resolved. The new peak in the red spectrum was consistent with the branch existing in the P configuration of the sample above the grey circle in Fig. \ref{cpw}(f). The microwave current applied to CPW1 with $f_{\rm irr}=4.3~$GHz hence led to the reversal of Py nanostripes at the remotely located CPW2. \\ \indent To investigate if heating of CPW1 by $i_{\rm rf}$ initiated the reversal we followed the same measurement protocol as applied in Fig. \ref{BLSoverview}(c) but changed the rf signal frequency to 1.25 GHz. The signal $i_{\rm rf}$ was applied for two hours, before taking the red spectrum in Fig. \ref{BLSoverview}(d). The red spectrum is found to contain the identical resonances as the black spectrum, i.e., $\mathbf{M}_{\rm Py}$ was not changed by applying an rf signal at 1.25~GHz. We explain the different spectra (red) in panels (c) and (d) of Fig. {\ref{BLSoverview} by the dispersion relation of YIG at 24 mT displayed in Fig. \ref{BLSoverview}(b). At $f_{\rm irr}=1.25~$GHz (green dashed line), magnons are not emitted into YIG as all allowed magnon bands reside at higher frequencies. This is different for $f_{\rm irr}=4.3~$GHz. Here, a Damon-Eshbach (DE) mode is allowed and excited at CPW1. It propagates to CPW2 as evidenced by the transmission spectra shown in Fig. \ref{cpw}. The allowed mode is the grating coupler mode $k_1+G$ with $\lambda=195$~nm. The characteristic resonance Y' in the red spectrum of Fig {\ref{BLSoverview}(c) evidences the reversal of Py nanostripes in Pos. 2.1 by the propagating magnon mode. In Pos. 2.2 located a few micrometers further away from CPW1 (Fig. \ref{1g25}a), we detected a modified spectrum (red) at 24 mT as well  (Fig. \ref{1g25}b). Hence, the reversal of Py nanostripes was induced in both gaps by the short-wave magnon mode $k_1+G$ excited at CPW1 at $f_{\rm irr}=4.3~$GHz.\\ \indent When applying a microwave signal with $f_{\rm irr}=5~$GHz to CPW1 (after again initializing sample A at -84 mT), we observed modified spectra (red) taken at Pos. 2.1 [Fig. \ref{1g25}(c)] and Pos. 2.2 [Fig. \ref{1g25}(d)]. Note that the directly excited grating coupler mode did not exist at 5 GHz. Still, the finding is different from the experiment conducted with $f_{\rm irr}=1.25~$GHz in Fig. \ref{BLSoverview}(d). We attribute the observed reversal of Py nanostripes to magnons which were excited by parametric pumping at CPW1 (Supplementary information). Their frequency reads $f_{\rm m} = f_{\rm irr}/2 = 2.5~$GHz, which was above the $k_1$ resonance ({$  f_{\rm k_1}$ = 2.3 GHz}) at 24 mT and inside the allowed magnon band for propagation. Such magnons hence reached CPW2 and could explain the observed reversal. \textcolor{blue}{We note that the phase-sensitive voltage detection of the VNA experiment does not allow us to evidence the magnons created by parametric pumping because of their shifted frequency. We resolve them by BLS as presented in detail in Fig. S6.}\\ \indent We performed experiments also at 2 mT after initializing the sample at -84 mT. In Pos. 2.1 [Fig. \ref{1g25}(e)] we observed that the magnon spectrum (red) was modified after applying $i_{\rm rf}$ with $f_{\rm irr}=1.25$~GHz. At 2 mT, spin waves at this small frequency were allowed [dotted lines in Fig. \ref{BLSoverview}(b)] and possessed a wave vector $k_1$  with $\lambda$ = 7222 nm. Excited at CPW1, the magnon mode changed the Py nanostripes underneath CPW2 at Pos. 2.1, but not at Pos. 2.2 [Fig. \ref{1g25}(f)]. We assume that at the small field the excitation of the $k_1$ mode \textcolor{blue}{was in the nonlinear regime as well, but additional parametric pumping did not take place at the small frequency. Instead, the  amplitude of magnons decayed due to enhanced scattering and was below the threshold for reversal in Pos. 2.2. The excitation of propagating magnons with a too high microwave power hence led to an incomplete reversal of nanostripes below CPW2. This finding is consistent with the non-monotonous variation of critical fields with applied microwave power reported in Ref. \cite{baumgaertl2023reversal} and as shown in Fig. \ref{switchingfield}.} \textcolor{blue}{BLS studies on magnon-induced reversal in sample B are shown in Fig. S7 and support the findings reported for sample A.}\\
\indent \textcolor{blue}{We now discuss the roles of the intermediate (spacer) layers.} The critical fields displayed in Fig. \ref{switchingfield} indicate that the insertion of the Cu spacer between the Py and YIG increased the switching field distribution and the coercivity of individual Py nanostripes compared to the SiO$_2$ spacer. Our spatially resolved BLS data demonstrate that the precessing magnetization of allowed spin waves in YIG creates a torque leading to an irreversible change of the nanostripe magnetization. The insertion of the Cu spacer excludes the transfer of exchange magnons as the main mechanism in contrast to assumptions made in Refs. \cite{Han1121,Wang2019b,Guo2021,Zheng2022}. Still, the Cu spacer thickness is smaller than the spin diffusion length \cite{SpinDiffLengthRev}. In this case, forced spin precession in YIG and concomitant spin pumping into Py might introduce an additional torque. \cite{Suresh2021} We noticed however significantly larger critical fields and a reduced switching efficiency at high power for sample A with Cu spacer compared to sample B with SiO$_2$ spacer. The data do not support the spin pumping effect to be relevant for reversal. Strikingly, we find the largest reduction in the critical fields $H_{\rm C1}$ and $H_{\rm C2}$ in sample B with the 5-nm-thick insulating spacer which avoids the spin-pumping torque. Thereby, we assume that dipolar coupling alone allowed for nanostripe reversal at a small power level. \\ \indent \textcolor{blue}{In the following we discuss the possible origin for the observed variation of spin-precessional power for magnon-induced reversal. We focus on Fig. \ref{criticalpowerstabrecap}(d), where we exclude direct microwave-assisted switching of nanostripes by the applied rf signal. In sample B, we observe the smallest spin-precessional power when the propagating magnons exhibit a wavelength of 195 nm underneath the gratings \cite{KBaumgaertl2020}. This value is (very close to) twice the width of a Py nanostripe. We argue that such a relation between the magnon wavelength and nanostripe width ensures the highest possible repetition rate by which a maximum in the dipolar stray field of a DE mode exerts a torque on the Py magnetization vector $\mathbf{M}_{\rm Py}$. For a shorter magnon wavelength a partial cancellation of the dynamic dipolar field occurs underneath a nanostripe, and the dynamic dipolar coupling is reduced. For a long wavelength the repetition rate is small by which the maxima of the dynamic stray field pass by the nanostripe and produce the relevant torque. These considerations motivate the observed minimum in Fig. \ref{criticalpowerstabrecap}(d) as a function of $f_{\rm irr}$, i.e., magnon wavelength. The slightly increased power levels needed for reversal in sample A incorporating the Cu spacer might indicate that additional spin pumping or an eddy current effect reduced the total torque. Further studies on stripes with different spacer layers and of e.g. different lengths and widths are needed to engineer their own eigenresonance frequency and explore in detail the hypothesis of a wavelength-dependent reversal mechanism drawn from the presented experiments.}   

\section*{Conclusions}
We reported magnon-induced reversal in Py/YIG hybrid structures with different intermediate layers. We quantified and compared the power values for magnon-induced  switching. Reversal of 100-nm-wide Py stripes was achieved by means of propagating magnons \textcolor{blue}{with wavelengths ranging from 148 nm to 7222 nm. Their excitation was realized both in the linear and non-linear regime. The non-linear parametric pumping was evidenced by local BLS microscopy.} {In an opposing field of 14 mT a spin-precessional power of \textcolor{blue}{the order of 1} nW was enough to reverse the up to 27 $\mu$m long Py nanostripes after magnon propagation over 15~$\mu$m. The absence of interlayer exchange coupling due to a spacer layer between Py and YIG led to nanostripes with partly enhanced coercive fields compared to Py stripes directly integrated on YIG. The enhanced coercive fields are advantageous in terms of a nonvolatile memory of magnon signals. Importantly, with increasing power, we achieved a reduction of switching fields of nanostripes to (nearly) zero mT}. Our results promise that \textcolor{blue}{nonvolatile magnon-signal storage in magnetic bits is feasible for wave-logic circuits and neural networks performing computational tasks at different magnon frequencies. Considerations based on dynamic dipolar coupling suggest that the power for magnon-induced storage might be minimized when the width of the magnetic bit equals half the wavelength of the magnon.}

\section*{Methods}
\textbf{A. Sample fabrication}. Devices are fabricated on 113-nm-thick YIG originating from the same wafer. The YIG had been deposited by liquid phase epitaxy on a 3-inch wafer and purchased by the company Matesy GmbH in Jena, Germany. The spacer is fabricated by DC sputtering of 5-nm-thick Cu on YIG. Then 20-nm-thick Py ($\rm Ni_{81}Fe_{19}$) is deposited via electron beam evaporation on the YIG. The gratings were written with electron beam lithography (EBL) using hydrogen silsesquioxane (HSQ) as negative resist and then transferred into the Py/Cu by ion beam etching. We etch both layers of Py and Cu. CPWs are fabricated via lift-off processing after EBL and Ti/Au (5 nm / 120 nm) evaporation. For sample B, the SiO$_2$ layer is deposited by e-beam evaporation and the following steps to fabricate stripes and CPW are unchanged.\\
\textbf{B. Broadband VNA spectra}. The broadband spectroscopy data $\Delta S_{\rm \alpha \beta}$ ($\rm \alpha ,  \, \beta$ = 1, 2) are obtained by nearest-neighbor subtraction of raw linear magnitude signals, i.e. $\Delta S_{\rm \alpha \beta}(f, H_{\rm i}) = S_{\rm \alpha \beta}(f,H_{\rm i+1})-S_{\rm \alpha \beta}(f,H_{\rm i})$. The magnetic field step is 2 mT. The linear magnitude signal Mag($S_{\rm \alpha \beta}$) is obtained from the quadrature sum of real (${{\rm Re}(S)}$) and imaginary (${{\rm Im}(S)}$) parts of median-subtracted signals. ${{\rm Re}(S)}$ (${{\rm Im}(S)}$) is obtained by the raw real (imaginary) part after removing at each measured frequency its median value across all applied magnetic fields.
\\
\textbf{C. Switching yield maps}. To build switching yield maps (Fig. \ref{switchyield}) we have followed the methodology described in Ref. \cite{baumgaertl2023reversal}. To get $\mu_0H_{\rm C}=14~$mT by the emission of the magnon mode $k_1$ ({in Ref. \cite{baumgaertl2023reversal}} this field $H_{\rm C}$ was labelled $H_{\rm C2}$.) We required ($-12 \pm 1$) dBm (63.1 $\mu$W) at CPW1. We compare this value to 58.4 $\mu$W needed {in Ref. \cite{baumgaertl2023reversal}}. The separation between CPW1 and CPW2 was larger by 20 $\mu$m  {in Ref. \cite{baumgaertl2023reversal}} compared to the present samples. However the previously reported critical field was only $+28$ mT compared to $+40$ mT in Fig. \ref{cpw}(d). This larger coercive field of nanostripes with the Cu underlayer used here might explain that a similar power level for reversal was needed though the propagation length of magnons was shorter.
\textcolor{blue}{To characterize the critical power levels featuring the switching at CPW1 (CPW2) we focus on the frequency branch 4.5$\div$4.7 (3.9$\div$4.1) GHz of the $S_{\rm 11}$ ($S_{\rm 21}$) spectra (cf. Fig. \ref{switchyield}).}
\\
\textcolor{blue}{To evaluate the critical precessional power $P_{\rm C,prec}$ we first extract the minimum critical power $P_{\rm C}$ for the same frequency. The overall irradiation frequency $f_{\rm irr}$ range is divided into sub-intervals of 250 MHz width. We record $P_{\rm C}$ and the frequency sub-interval $\delta f_{\rm P_{\rm C}}$ that achieves $P_{\rm C}$. These measurements are conducted with 1 kHz bandwidth and 250 MHz frequency resolution. Examples of such measurements are reported in Fig. \ref{switchyield}. The magnetic field is 14 mT.
To obtain the relevant Mag(S$_{\rm 11}$) signal we acquire field-dependent reflection spectra with 0.1 kHz bandwidth and 3.3 MHz frequency resolution. The VNA power for these measurements is labelled $P_{\rm b}$. $P_{\rm b}$ equals -25 (-10) dBm for datasets that are analysed to evaluate $P_{\rm C1,prec}$ ($P_{\rm C2,prec}$). The magnetic field is swept from -90 mT to positive fields larger than $\mu_{\rm 0}H_{\rm C2}$. With these datasets we define for both real and imaginary parts a median value across all applied magnetic fields at each frequency point. Then the linear magnitude signal  Mag($S_{\rm 11}$) is constructed as described in paragraph B of the Methods section. We focus on the frequency range defined by the previously found $\delta f_{\rm P_{\rm C}}$ and consider the reflection spectrum in the same range. Inside this frequency range we identify the frequency value $f_{\rm *}$ that achieves the local maximum of Mag($S_{\rm 11}$): $(\text{Mag}(S_{\rm 11}))_{\rm *}$. This represents the maximum absorbed energy by the spin system. The critical spin precessional power is then evaluated by $P_{\rm C,prec}$ = $P_{\rm C} \cdot[(\text{Mag}(S_{\rm 11}))_{\rm *}]^2$}. 
\\
\textbf{D. BLS measurement protocol}. To acquire the BLS spectra we initialize the system by applying -82 mT with a permanent magnet we then gradually increased the field to reach the targeted positive value. In so doing the system reaches the AP state. Thermal magnon spectra are acquired before injection of any rf signal at CPW1. We apply rf signal at CPW1 at a fixed frequency for increasing nominal powers. At each power step, we record the BLS signal while having the rf on. The rf irradiation at each power level is approximately 2 hours long. At the end of the experiment, after switching off the rf generator, the thermal magnon spectra is measured again and compared to the one acquired in the 'as-prepared' AP state. To minimize spatial drift and maintain the same position of the laser spot we used a feedback system with image recognition acting every 5 minutes. For the BLS experiments the sample is wire-bonded to a PCB. The power levels that we discussed for BLS measurements are meant as nominal values.\\

\begin{suppinfo}
Numerical evaluation of the excitation spectrum of the CPW inhomogeneous dynamic field conducted by combining COMSOL simulation and FFT analysis.

Broadband reflection spin wave spectra at $P_{\rm VNA}$ = -25 dBm.

Protocol for evaluation of nearly zero critical fields for nanomagnet reversal.

Experimental datasets acquired with Brillouin light scattering microscopy ($\mu$BLS) at the emitter CPW (CPW1) investigating magnon-induced magnetization reversal of the nanomagnets beneath CPW1.

$\mu$BLS datasets of the Py nanostripes during continous-wave excitation, at different power levels, of multiple magnon modes in the underlying YIG.

$\mu$BLS experiments reporting magnon-induced magnetization reversal for another device with hybrid interface Py/SiO$_{\rm 2}$/YIG.

Inductive broadband spectroscopy measurements at -25 dBm acquired for samples A, C and B for comparison of reflection and transmission spectra.
\end{suppinfo}

\section*{Acknowledgments}
{The authors have used the colour maps for visualization of the VNA data provided by Fabio Crameri. The Scientific colour map bam \cite{crameri2018scientific} is used in this study to prevent visual distortion of the data and exclusion of readers with colour-vision deficiencies \cite{crameri2020misuse}.}
The authors acknowledge experimental support by Ping Che and discussions with Shreyas Joglekar and Mohammad Hamdi.

\subsection*{Funding}
The research was supported by the SNSF via grant number 197360.

\subsection*{Author contributions}
D.G., K.B. and A.M. planned the experiments and designed the samples. A.M. prepared the samples and performed the experiments together with K.B. A.M. and D.G. analyzed and interpreted the data. A.M. and D.G. wrote the manuscript. All authors commented on the manuscript.

\subsection*{Competing interests}
The authors declare that they have no competing interests.

\subsection*{Data availability}
The datasets generated and/or analysed during the current study are available from the corresponding author on reasonable request.

\subsection*{Correspondence}
Correspondence and requests for materials should be addressed to D.G.(email: dirk.grundler@epfl.ch).\\

\bibliography{CorrectedBibliography}

\providecommand{\latin}[1]{#1}
\makeatletter
\providecommand{\doi}
  {\begingroup\let\do\@makeother\dospecials
  \catcode`\{=1 \catcode`\}=2 \doi@aux}
\providecommand{\doi@aux}[1]{\endgroup\texttt{#1}}
\makeatother
\providecommand*\mcitethebibliography{\thebibliography}
\csname @ifundefined\endcsname{endmcitethebibliography}
  {\let\endmcitethebibliography\endthebibliography}{}
\begin{mcitethebibliography}{27}
\providecommand*\natexlab[1]{#1}
\providecommand*\mciteSetBstSublistMode[1]{}
\providecommand*\mciteSetBstMaxWidthForm[2]{}
\providecommand*\mciteBstWouldAddEndPuncttrue
  {\def\EndOfBibitem{\unskip.}}
\providecommand*\mciteBstWouldAddEndPunctfalse
  {\let\EndOfBibitem\relax}
\providecommand*\mciteSetBstMidEndSepPunct[3]{}
\providecommand*\mciteSetBstSublistLabelBeginEnd[3]{}
\providecommand*\EndOfBibitem{}
\mciteSetBstSublistMode{f}
\mciteSetBstMaxWidthForm{subitem}{(\alph{mcitesubitemcount})}
\mciteSetBstSublistLabelBeginEnd
  {\mcitemaxwidthsubitemform\space}
  {\relax}
  {\relax}

\bibitem[Khitun \latin{et~al.}(2010)Khitun, Bao, and Wang]{Khitun_2010}
Khitun,~A.; Bao,~M.; Wang,~K.~L. Magnonic logic circuits. \emph{J. Phys. D:
  Appl. Phys.} \textbf{2010}, \emph{43}, 264005\relax
\mciteBstWouldAddEndPuncttrue
\mciteSetBstMidEndSepPunct{\mcitedefaultmidpunct}
{\mcitedefaultendpunct}{\mcitedefaultseppunct}\relax
\EndOfBibitem
\bibitem[Chumak \latin{et~al.}(2014)Chumak, Serga, and
  Hillebrands]{ChumakMagTrans}
Chumak,~A.; Serga,~A.; Hillebrands,~B. Magnon transistor for all-magnon data
  processing. \emph{Nat. Commun.} \textbf{2014}, \emph{5}, 4700\relax
\mciteBstWouldAddEndPuncttrue
\mciteSetBstMidEndSepPunct{\mcitedefaultmidpunct}
{\mcitedefaultendpunct}{\mcitedefaultseppunct}\relax
\EndOfBibitem
\bibitem[Mahmoud \latin{et~al.}(2020)Mahmoud, Ciubotaru, Vanderveken, Chumak,
  Hamdioui, Adelmann, and Cotofana]{Mahmoud2020}
Mahmoud,~A.; Ciubotaru,~F.; Vanderveken,~F.; Chumak,~A.~V.; Hamdioui,~S.;
  Adelmann,~C.; Cotofana,~S. Introduction to spin wave computing. \emph{J.
  Appl. Phys.} \textbf{2020}, \emph{128}, 161101\relax
\mciteBstWouldAddEndPuncttrue
\mciteSetBstMidEndSepPunct{\mcitedefaultmidpunct}
{\mcitedefaultendpunct}{\mcitedefaultseppunct}\relax
\EndOfBibitem
\bibitem[Yu \latin{et~al.}(2013)Yu, Duerr, Huber, Bahr, Schwarze, Brandl, and
  Grundler]{Yu2013}
Yu,~H.; Duerr,~G.; Huber,~R.; Bahr,~M.; Schwarze,~T.; Brandl,~F.; Grundler,~D.
  Omnidirectional spin-wave nanograting coupler. \emph{Nat. Commun.}
  \textbf{2013}, \emph{4}, 2702\relax
\mciteBstWouldAddEndPuncttrue
\mciteSetBstMidEndSepPunct{\mcitedefaultmidpunct}
{\mcitedefaultendpunct}{\mcitedefaultseppunct}\relax
\EndOfBibitem
\bibitem[Yu \latin{et~al.}(2016)Yu, Kelly, Cros, Bernard, Bortolotti, Anane,
  Brandl, Heimbach, and Grundler]{yu2016approaching}
Yu,~H.; Kelly,~O.~d.; Cros,~V.; Bernard,~R.; Bortolotti,~P.; Anane,~A.;
  Brandl,~F.; Heimbach,~F.; Grundler,~D. Approaching soft X-ray wavelengths in
  nanomagnet-based microwave technology. \emph{Nat. Commun.} \textbf{2016},
  \emph{7}, 11255\relax
\mciteBstWouldAddEndPuncttrue
\mciteSetBstMidEndSepPunct{\mcitedefaultmidpunct}
{\mcitedefaultendpunct}{\mcitedefaultseppunct}\relax
\EndOfBibitem
\bibitem[Chen \latin{et~al.}(2019)Chen, Yu, Liu, Liu, Madami, Shen, Zhang, Tu,
  Alam, Xia, Wu, Gubbiotti, Blanter, Bauer, and Yu]{PhysRevB.100.104427}
Chen,~J.; Yu,~T.; Liu,~C.; Liu,~T.; Madami,~M.; Shen,~K.; Zhang,~J.; Tu,~S.;
  Alam,~M.~S.; Xia,~K.; Wu,~M.; Gubbiotti,~G.; Blanter,~Y.~M.; Bauer,~G. E.~W.;
  Yu,~H. Excitation of unidirectional exchange spin waves by a nanoscale
  magnetic grating. \emph{Phys. Rev. B} \textbf{2019}, \emph{100}, 104427\relax
\mciteBstWouldAddEndPuncttrue
\mciteSetBstMidEndSepPunct{\mcitedefaultmidpunct}
{\mcitedefaultendpunct}{\mcitedefaultseppunct}\relax
\EndOfBibitem
\bibitem[Baumgaertl \latin{et~al.}(2020)Baumgaertl, Gräfe, Che, Mucchietto,
  Förster, Träger, Bechtel, Weigand, Schütz, and Grundler]{KBaumgaertl2020}
Baumgaertl,~K.; Gräfe,~J.; Che,~P.; Mucchietto,~A.; Förster,~J.; Träger,~N.;
  Bechtel,~M.; Weigand,~M.; Schütz,~G.; Grundler,~D. Nanoimaging of Ultrashort
  Magnon Emission by Ferromagnetic Grating Couplers at {GH}z Frequencies.
  \emph{Nano Lett.} \textbf{2020}, \emph{20}, 7281--7286, PMID: 32830984\relax
\mciteBstWouldAddEndPuncttrue
\mciteSetBstMidEndSepPunct{\mcitedefaultmidpunct}
{\mcitedefaultendpunct}{\mcitedefaultseppunct}\relax
\EndOfBibitem
\bibitem[Liu \latin{et~al.}(2018)Liu, Chen, Liu, Heimbach, Yu, Xiao, Hu, Liu,
  Chang, Stueckler, Tu, Zhang, Zhang, Gao, Liao, Yu, Xia, Lei, Zhao, and
  Wu]{Liu2018}
Liu,~C. \latin{et~al.}  Long-distance propagation of short-wavelength spin
  waves. \emph{Nat. Commun.} \textbf{2018}, \emph{9}, 738\relax
\mciteBstWouldAddEndPuncttrue
\mciteSetBstMidEndSepPunct{\mcitedefaultmidpunct}
{\mcitedefaultendpunct}{\mcitedefaultseppunct}\relax
\EndOfBibitem
\bibitem[Watanabe \latin{et~al.}(2023)Watanabe, Bhat, Mucchietto, Dayi, Shan,
  and Grundler]{SWatanabe2023}
Watanabe,~S.; Bhat,~V.~S.; Mucchietto,~A.; Dayi,~E.~N.; Shan,~S.; Grundler,~D.
  Periodic and Aperiodic NiFe Nanomagnet/Ferrimagnet Hybrid Structures for 2D
  Magnon Steering and Interferometry with High Extinction Ratio. \emph{Adv.
  Mater.} \textbf{2023}, \relax
\mciteBstWouldAddEndPunctfalse
\mciteSetBstMidEndSepPunct{\mcitedefaultmidpunct}
{}{\mcitedefaultseppunct}\relax
\EndOfBibitem
\bibitem[Wang \latin{et~al.}(2021)Wang, Madami, Chen, Sheng, Zhao, Zhang, He,
  Guo, Jia, Liu, Song, Han, Yu, Gubbiotti, and Yu]{ChiralPumpSW}
Wang,~H.; Madami,~M.; Chen,~J.; Sheng,~L.; Zhao,~M.; Zhang,~Y.; He,~W.;
  Guo,~C.; Jia,~H.; Liu,~S.; Song,~Q.; Han,~X.; Yu,~D.; Gubbiotti,~G.; Yu,~H.
  Tunable Damping in Magnetic Nanowires Induced by Chiral Pumping of Spin
  Waves. \emph{ACS Nano} \textbf{2021}, \emph{15}, 9076--9083, PMID:
  33977721\relax
\mciteBstWouldAddEndPuncttrue
\mciteSetBstMidEndSepPunct{\mcitedefaultmidpunct}
{\mcitedefaultendpunct}{\mcitedefaultseppunct}\relax
\EndOfBibitem
\bibitem[Baumgaertl and Grundler(2023)Baumgaertl, and
  Grundler]{baumgaertl2023reversal}
Baumgaertl,~K.; Grundler,~D. Reversal of nanomagnets by propagating magnons in
  ferrimagnetic yttrium iron garnet enabling nonvolatile magnon memory.
  \emph{Nature Communications} \textbf{2023}, \emph{14}, 1490\relax
\mciteBstWouldAddEndPuncttrue
\mciteSetBstMidEndSepPunct{\mcitedefaultmidpunct}
{\mcitedefaultendpunct}{\mcitedefaultseppunct}\relax
\EndOfBibitem
\bibitem[Islam \latin{et~al.}(2019)Islam, Li, Chen, Wan, Chen, Gao, Wu, Yu,
  Saraswat, and Wong]{Islam_2019}
Islam,~R.; Li,~H.; Chen,~P.-Y.; Wan,~W.; Chen,~H.-Y.; Gao,~B.; Wu,~H.; Yu,~S.;
  Saraswat,~K.; Wong,~H.-S.~P. Device and materials requirements for
  neuromorphic computing. \emph{J. Phys. D: Appl. Phys.} \textbf{2019},
  \emph{52}, 113001\relax
\mciteBstWouldAddEndPuncttrue
\mciteSetBstMidEndSepPunct{\mcitedefaultmidpunct}
{\mcitedefaultendpunct}{\mcitedefaultseppunct}\relax
\EndOfBibitem
\bibitem[Sebastian \latin{et~al.}(2020)Sebastian, Gallo, Khaddam-Aljameh, and
  Eleftheriou]{Sebastian2020}
Sebastian,~A.; Gallo,~M.~L.; Khaddam-Aljameh,~R.; Eleftheriou,~E. Memory
  devices and applications for in-memory computing. \emph{Nat. Nanotechn.}
  \textbf{2020}, \emph{15}, 529--544\relax
\mciteBstWouldAddEndPuncttrue
\mciteSetBstMidEndSepPunct{\mcitedefaultmidpunct}
{\mcitedefaultendpunct}{\mcitedefaultseppunct}\relax
\EndOfBibitem
\bibitem[Papp \latin{et~al.}(2021)Papp, Porod, and Csaba]{Papp2021}
Papp,~{\'A}.; Porod,~W.; Csaba,~G. Nanoscale neural network using non-linear
  spin-wave interference. \emph{Nat. Commun.} \textbf{2021}, \emph{12},
  6422\relax
\mciteBstWouldAddEndPuncttrue
\mciteSetBstMidEndSepPunct{\mcitedefaultmidpunct}
{\mcitedefaultendpunct}{\mcitedefaultseppunct}\relax
\EndOfBibitem
\bibitem[Klingler \latin{et~al.}(2018)Klingler, Amin, Gepr\"ags, Ganzhorn,
  Maier-Flaig, Althammer, Huebl, Gross, McMichael, Stiles, Goennenwein, and
  Weiler]{Klingler2018}
Klingler,~S.; Amin,~V.; Gepr\"ags,~S.; Ganzhorn,~K.; Maier-Flaig,~H.;
  Althammer,~M.; Huebl,~H.; Gross,~R.; McMichael,~R.~D.; Stiles,~M.~D.;
  Goennenwein,~S. T.~B.; Weiler,~M. Spin-Torque Excitation of Perpendicular
  Standing Spin Waves in Coupled $\mathrm{YIG}/\mathrm{Co}$ Heterostructures.
  \emph{Phys. Rev. Lett.} \textbf{2018}, \emph{120}, 127201\relax
\mciteBstWouldAddEndPuncttrue
\mciteSetBstMidEndSepPunct{\mcitedefaultmidpunct}
{\mcitedefaultendpunct}{\mcitedefaultseppunct}\relax
\EndOfBibitem
\bibitem[Maendl and Grundler(2018)Maendl, and Grundler]{Maendl2018}
Maendl,~S.; Grundler,~D. Multi-directional emission and detection of spin waves
  propagating in yttrium iron garnet with wavelengths down to about 100 nm.
  \emph{Appl. Phys. Lett.} \textbf{2018}, \emph{112}, 192410\relax
\mciteBstWouldAddEndPuncttrue
\mciteSetBstMidEndSepPunct{\mcitedefaultmidpunct}
{\mcitedefaultendpunct}{\mcitedefaultseppunct}\relax
\EndOfBibitem
\bibitem[Watanabe \latin{et~al.}(2021)Watanabe, Bhat, Baumgaertl, Hamdi, and
  Grundler]{Watanabe2021}
Watanabe,~S.; Bhat,~V.; Baumgaertl,~K.; Hamdi,~M.; Grundler,~D. Direct
  observation of multiband transport in magnonic Penrose quasicrystals via
  broadband and phase-resolved spectroscopy. \emph{Sci. Adv.} \textbf{2021},
  \emph{7}, eabg3771\relax
\mciteBstWouldAddEndPuncttrue
\mciteSetBstMidEndSepPunct{\mcitedefaultmidpunct}
{\mcitedefaultendpunct}{\mcitedefaultseppunct}\relax
\EndOfBibitem
\bibitem[Bass and Pratt(2007)Bass, and Pratt]{SpinDiffLengthRev}
Bass,~J.; Pratt,~W.~P. Spin-diffusion lengths in metals and alloys, and
  spin-flipping at metal/metal interfaces: an experimentalist's critical
  review. \emph{Journal of Physics: Condensed Matter} \textbf{2007}, \emph{19},
  183201\relax
\mciteBstWouldAddEndPuncttrue
\mciteSetBstMidEndSepPunct{\mcitedefaultmidpunct}
{\mcitedefaultendpunct}{\mcitedefaultseppunct}\relax
\EndOfBibitem
\bibitem[Suresh \latin{et~al.}(2021)Suresh, Bajpai,
  Petrovi\ifmmode~\acute{c}\else \'{c}\fi{}, Yang, and
  Nikoli\ifmmode~\acute{c}\else \'{c}\fi{}]{Suresh2021}
Suresh,~A.; Bajpai,~U.; Petrovi\ifmmode~\acute{c}\else \'{c}\fi{},~M.~D.;
  Yang,~H.; Nikoli\ifmmode~\acute{c}\else \'{c}\fi{},~B.~K. Magnon- versus
  Electron-Mediated Spin-Transfer Torque Exerted by Spin Current across an
  Antiferromagnetic Insulator to Switch the Magnetization of an Adjacent
  Ferromagnetic Metal. \emph{Phys. Rev. Applied} \textbf{2021}, \emph{15},
  034089\relax
\mciteBstWouldAddEndPuncttrue
\mciteSetBstMidEndSepPunct{\mcitedefaultmidpunct}
{\mcitedefaultendpunct}{\mcitedefaultseppunct}\relax
\EndOfBibitem
\bibitem[Gurevich and Melkov(1996)Gurevich, and Melkov]{Gurevich96}
Gurevich,~A.~G.; Melkov,~G.~A. \emph{Magnetization oscillations and waves}; CRC
  Press, Boca Raton, 1996\relax
\mciteBstWouldAddEndPuncttrue
\mciteSetBstMidEndSepPunct{\mcitedefaultmidpunct}
{\mcitedefaultendpunct}{\mcitedefaultseppunct}\relax
\EndOfBibitem
\bibitem[Han \latin{et~al.}(2019)Han, Zhang, Hou, Siddiqui, and Liu]{Han1121}
Han,~J.; Zhang,~P.; Hou,~J.~T.; Siddiqui,~S.~A.; Liu,~L. Mutual control of
  coherent spin waves and magnetic domain walls in a magnonic device.
  \emph{Science} \textbf{2019}, \emph{366}, 1121--1125\relax
\mciteBstWouldAddEndPuncttrue
\mciteSetBstMidEndSepPunct{\mcitedefaultmidpunct}
{\mcitedefaultendpunct}{\mcitedefaultseppunct}\relax
\EndOfBibitem
\bibitem[Wang \latin{et~al.}(2019)Wang, Zhu, Yang, Lee, Mishra, Go, Oh, Kim,
  Cai, Liu, Pollard, Shi, Lee, Teo, Wu, Lee, and Yang]{Wang2019b}
Wang,~Y. \latin{et~al.}  Magnetization switching by magnon-mediated spin torque
  through an antiferromagnetic insulator. \emph{Science} \textbf{2019},
  \emph{366}, 1125--1128\relax
\mciteBstWouldAddEndPuncttrue
\mciteSetBstMidEndSepPunct{\mcitedefaultmidpunct}
{\mcitedefaultendpunct}{\mcitedefaultseppunct}\relax
\EndOfBibitem
\bibitem[Guo \latin{et~al.}(2021)Guo, Wan, Zhao, Fang, Ma, Wang, Yan, He, Xing,
  Feng, and Han]{Guo2021}
Guo,~C.~Y.; Wan,~C.~H.; Zhao,~M.~K.; Fang,~C.; Ma,~T.~Y.; Wang,~X.; Yan,~Z.~R.;
  He,~W.~Q.; Xing,~Y.~W.; Feng,~J.~F.; Han,~X.~F. Switching the perpendicular
  magnetization of a magnetic insulator by magnon transfer torque. \emph{Phys.
  Rev. B} \textbf{2021}, \emph{104}, 094412\relax
\mciteBstWouldAddEndPuncttrue
\mciteSetBstMidEndSepPunct{\mcitedefaultmidpunct}
{\mcitedefaultendpunct}{\mcitedefaultseppunct}\relax
\EndOfBibitem
\bibitem[Zheng \latin{et~al.}(2022)Zheng, Lan, Fang, Li, Liu, Ledesma-Martin,
  Wen, Li, Zhang, Ma, Qiu, Liu, Manchon, and Zhang]{Zheng2022}
Zheng,~D.; Lan,~J.; Fang,~B.; Li,~Y.; Liu,~C.; Ledesma-Martin,~J.~O.; Wen,~Y.;
  Li,~P.; Zhang,~C.; Ma,~Y.; Qiu,~Z.; Liu,~K.; Manchon,~A.; Zhang,~X.
  High-Efficiency Magnon-Mediated Magnetization Switching in All-Oxide
  Heterostructures with Perpendicular Magnetic Anisotropy. \emph{Advanced
  Materials} \textbf{2022}, 2203038\relax
\mciteBstWouldAddEndPuncttrue
\mciteSetBstMidEndSepPunct{\mcitedefaultmidpunct}
{\mcitedefaultendpunct}{\mcitedefaultseppunct}\relax
\EndOfBibitem
\bibitem[Crameri(2018)]{crameri2018scientific}
Crameri,~F. Scientific colour maps. \emph{Zenodo} \textbf{2018},
  \emph{10}\relax
\mciteBstWouldAddEndPuncttrue
\mciteSetBstMidEndSepPunct{\mcitedefaultmidpunct}
{\mcitedefaultendpunct}{\mcitedefaultseppunct}\relax
\EndOfBibitem
\bibitem[Crameri \latin{et~al.}(2020)Crameri, Shephard, and
  Heron]{crameri2020misuse}
Crameri,~F.; Shephard,~G.~E.; Heron,~P.~J. The misuse of colour in science
  communication. \emph{Nature communications} \textbf{2020}, \emph{11},
  5444\relax
\mciteBstWouldAddEndPuncttrue
\mciteSetBstMidEndSepPunct{\mcitedefaultmidpunct}
{\mcitedefaultendpunct}{\mcitedefaultseppunct}\relax
\EndOfBibitem
\end{mcitethebibliography}

\end{document}